# Implicit Recursive Characteristics of STOP

Michael H. Ji
Laboratory for Foundations of Computer Science
School of Informatics
University of Edinburgh, UK
mji@ed.ac.uk

**Abstract:**
The most important notations of Communicating Sequential Process(CSP) are the *process* and the *prefix (event)→(process)* operator. While we can formally apply the → operator to define a live process's behavior, the STOP process, which usually resulted from deadlock, starving or livelock, is lack of formal description, defined by most literatures as "doing nothing but halt". In this paper, we argue that the STOP process should not be considered as a black box, it should follow the prefix → schema and the same inference rules so that a unified and consistent process algebra model can be established. In order to achieve this goal, we introduce a special event called "nil" that any process can take. This nil event will do nothing meaningful and leave nothing on a process's observable record. With the nil event and its well-defined rules, we can successfully use the → operator to formally describe a process's complete behavior in its whole life circle. More interestingly, we can use prefix → and nil event to fully describe the STOP process's internal behavior and conclude that the STOP's formal equation can be given as simple as $STOP_{\alpha X} = \mu X.\ nil \rightarrow X$.

**Keywords:** Communicating Sequential Process, Process, STOP, SKIP, Trace

## 1. Introduction

In a now classic book, published in 1985, Turing Award winner Dr. Tony Hoare proposed Communicating Sequential Process(CSP) [1]. The CSP book introduced the notion of *process* and the *prefix operator (event)→(process)*, intended as a mathematical abstraction of the interactions between a computing system and its environment [2]. A process is usually categorized as either a live process or a STOP process. A live process can interact with its environment while a STOP process is considered as a breaking process [1,2]. For example:

- ($coin \rightarrow choc \rightarrow STOP_{\alpha VMS}$) a simple vending machine which consumes one coin before breaking.

The "breaking" semantic in CSP literatures is defined as a scenario that a process falls through deadlock, starving, or live-lock situations and thus a process cannot any more engage in events from the environment. For example, a vending machine runs out of chocolate so it cannot respond to new coin event. While we can formally apply the (event)→(process) to define a live process's behavior, the STOP process is lack of formal description, defined by most literatures as a "doing nothing but halt" black box [1,2]. However, as we know, those breaking processes are not physically dead yet; they are just in some special statuses that could not respond to outside events any more until those stalled situations get changed.
In this paper, we argue that the STOP process should not be considered as a black box, and should follow the prefix → schema and the same inference rules so that a unified and consistent process algebra model can be established. In order to achieve this goal, we introduce a special event called "nil" that any process can take. This nil event will do nothing meaningful and leave nothing on a process's observable record. With the nil event and its well-defined rules, we can successfully use the → operator to formally describe a process's complete behavior in its whole life circle. More interestingly, we can use prefix → and nil event to fully describe the STOP process's internal behavior and conclude that the STOP's formal equation can be given as simple as $STOP_{\alpha X} = \mu$ X. nil →X.

## 2. Unified Process Model

In CSP, a process is defined recursively. The → operator always takes an event on the left and a process on the right.

<process> = <event> → <process>

In other words, a process is defined by an event and a followed process. The followed process itself follows the same → formula. Let x be an event and let Q be a process, (x → Q) describes a process P which first engages in the event x and then behaves exactly as described by Q.

P = (x → Q)

Since Q itself is a process, it must also follow the prefix operator definition. Assume that Q = ( y → R ), then, P can be expanded as:

P = (x → y → R)

Assuming that $x_0$, $x_1$,……, $x_{n-1}$ are the event sequences when the process P interacts with its environment, then P's behavior will look like:

P = ($x_0$ → $x_1$ → … → $x_{n-1}$ → $P_n$), where $P_n$ is a live or a STOP process[1].

Now, we investigate how a STOP process originates. We can use take the VMS process as an example [1]. Assume this vending machine initially has only 2 chocolates, the VMS's behavior can be depicted as below:

VMS = (coin → (choc → (coin → (choc → $STOP_{\alpha VMS}$))))

Note that the VMS and the $STOP_{\alpha VMS}$ have the same alphabet(events) that is αVMS = αSTOP = {coin, choc}.
In order to establish the unified process model based on the → operator, we need investigate what the inside is for the black box $STOP_{\alpha VMS}$. Suppose $STOP_{\alpha VMS}$ can also be depicted with the → schema as the VMS's first four actions (coin, choc, coin, choc): $STOP_{\alpha VMS}$ = (x → P), then, VMS can be expanded one more step further as:

VMS = (coin → (choc → (coin → (choc → (x → P)))))

Because VMS reaches $STOP_{\alpha VMS}$ stage when the VMS runs out of its two chocolates, thus the $STOP_{\alpha VMS}$ process cannot respond to any more coin unless it gets refilled. Therefore, the x of “(x → P)” cannot not be any event from αSTOP alphabet. Therefore, with {coin, choc} only as αSTOP’s alphabet, prefix (x→ P) formula cannot be adopted for $STOP_{\alpha VMS}$. We have to treat $STOP_{\alpha VMS}$ as a zombie, e.g., “doing nothing but halt” black box [1].

---

1 We don't consider SKIP process here. SKIP and STOP have similar characteristics, except the last event that being executed.

In order to solve this above dilemma, we create a special event for any process: A “nil” event notation is introduced to express the nothing or empty event “ ”, and by default, belongs to any process’s alphabet. We define that the nil event has the following characteristics:

- nil event can be executed instantly[2] by any process without printing any record on the trace that can be observed from the environment.
- nil event satisfies the below important laws and rules, alongside all other default laws and rules from CSP theory [1,2].

**Laws for nil**

**L1** (nil → (x → P)) = (x → P)
The process will instantly move to process event x. (by definition)
**L2** (x → (nil → P)) = (x → P)
The process will, after event x, instantly start to process the first event of P. (by definition)
**L3** (nil → P) = P (by L1)
**L4** (nil → P) ≠ STOP
Proof: If P = (x → Q), then (nil → P) = (nil → x → Q) = (x → Q) ≠ STOP (by L1)
**L5** (nil → P) || (nil → Q) = (P || Q))
**Proof:**
Based on CSP 2.3.1 L4A, nil is an event for both P and Q
then
(nil → P) || (nil → Q) = (nil → (P || Q))
then
(nil → (P || Q)) = (P || Q)) (by L1)
**L6** (nil → P) || (x → Q) = (P || (x →Q)) (by L3)

**Operations on Traces for nil**

Based on definition, nil event will not leave any record on a process’s trace and satisfies below set of rules.

**L1** <nil> = < >
nil event introduces nothing on process’s trace record.
**L2** <nil*> = < >
start of nil events is the same as one single nil event.
**L3** <x><nil> = <x>
nil event does not affect its previous trace.
**L4** <nil><x> = <x>
nil event does not affect its afterward trace.
**L5** <x><nil><y> = <x, y>
nil event in the middle of two traces can be omitted.

---

2 “instantly” means that a process will cost nothing (time, space) for executing a nil event.

With the definition of nil event and being introduced into a process's default alphabet, we can revisit our above VMS example. First, we extend $STOP_{\alpha VMS}$ alphabet as $\alpha VMS = \alpha VMS \cup \{nil\} = \{coin, choc, nil\}$; then we start to prove that nil event can be seamlessly used to formally describe VMS's last stage's $STOP_{\alpha VMS}$ behavior.
**Proof**:

For $STOP_{\alpha VMS} = (x \rightarrow P)$, let x = nil, then, $STOP_{\alpha VMS} = (nil \rightarrow P)$

then VMS is expanded as:

$$VMS = (coin \rightarrow (choc \rightarrow (coin \rightarrow (choc \rightarrow nil \rightarrow P))))$$

From an observer's viewpoint, the above VMS's observable trace is still <coin, choc, coin, choc>, satisfies the VMS's specification[3]. The proof is straightforward: <coin, choc, coin, choc, nil> = <coin, choc, coin, choc, nil> (by nil's trace L3).

Since P is also a process, suppose

$$P_{\alpha VMS} = (y \rightarrow Q)$$

Therefore, if this VMS keeps expanding, its behavior is:

$$VMS = (coin \rightarrow (choc \rightarrow (coin \rightarrow (choc \rightarrow nil \rightarrow (y \rightarrow Q)))))$$

then

$$VMS = (coin \rightarrow (choc \rightarrow (coin \rightarrow (choc \rightarrow (y \rightarrow Q))))) \text{ (by L3)}$$

Apparently, y must be nil event in order to keep VMS's specification and STOP semantics after running out of two chocolates. Then, we have:

$$VMS = (coin \rightarrow (choc \rightarrow (coin \rightarrow (choc \rightarrow nil \rightarrow (nil \rightarrow Q)))))$$

Hence, if we keep expanding the $STOP_{\alpha VMS}$ right side, we can easily prove that the $STOP_{\alpha VMS}$ process can be rewritten as follows:

$$STOP_{\alpha VMS} = (nil \rightarrow nil \rightarrow nil \rightarrow nil \rightarrow nil \rightarrow nil \rightarrow nil \rightarrow \ldots )$$

and

$$VMS = (coin \rightarrow (choc \rightarrow (coin \rightarrow (choc \rightarrow (nil \rightarrow (nil \rightarrow (nil \rightarrow (nil \rightarrow \ldots))))))))$$

In other words, nil event will be the only event that $STOP_{\alpha VMS}$ can be executed until the VMS machine get refilled. Its recursive equation can be concluded as below:

$$\mathbf{STOP_{\alpha VMS} = \mu\, X\, .\, nil \rightarrow X}$$
or

[3] In CSP, specification is a function description. For example, "A vender machine that will first output two chocolates and then STOP"

$\mathbf{STOP_{\alpha VMS} = (nil \rightarrow STOP_{\alpha VMS})}$

Up to here, we can successfully establish a unified algebra model for VMS. And more importantly, it is only prefix → based:

VMS = (coin → (choc → (coin → (choc → μ X . nil → X))))

Based on nil event's definition, law and rules, we can easily prove that the above VMS's equation satisfies well the VMS's specification. Its observable trace will be still <coin, choc, coin, choc>.

## 3. Discussions

In CSP, in order to describe a sequential process, a successful termination event ✓ (pronounced "success") is introduced. A sequential process is defined as that the last event that the process executes is ✓ before doing nothing. Correspondingly, The SKIP process is defined as a process which does nothing but terminate successfully. For example, A vending machine that is intended to serve only one customer with chocolate or toffee and then terminate successfully [1].

VMONE = (coin → (choc → SKIP | toffee → SKIP ))

When we investigate the SKIP process in more detail, a similar question arises: What's the SKIP's formal description with → operator? Assume SKIP = (✓ → P), then what P would be? It cannot be a STOP process, which semantics is for deadlock, starving or live-lock issues. If SKIP = (✓ → STOP), it means: SKIP = (✓ → μ X . nil → X ). Then this breaks SKIP's definition. The last event being executed becomes nil, instead of ✓. In other words, a process cannot get successful terminated and then broken. We propose that SKIP's formal equation should be also a recursive process as below. And the proof is very straightforward:

**SKIP = μ X . ✓ → X**
**or**
**SKIP = (✓ → SKIP)**

The VMONE equation can be rewritten as:

VMONE = (coin → (choc → μ X . ✓ → X | toffee → μ X . ✓ → X))

**Acknowledgments**

I want to thank professor Philip Wadler for his review and insightful comments.